\documentclass[conference]{IEEEtran}

\usepackage{array}
\newcolumntype{P}[1]{>{\centering\arraybackslash}p{#1}}
\usepackage{color}
\usepackage{multirow}
\usepackage{amsmath}
\usepackage{algorithm}
\usepackage[noend]{algpseudocode}
\usepackage{mathtools}
\usepackage{svg}
\usepackage{amsmath}

 \usepackage{graphicx}

\usepackage{cite}

\begin{document}

\title{Optimized Implementation of Neuromorphic HATS Algorithm on FPGA}
\author{\IEEEauthorblockN{Khushal Sethi, Manan Suri}}

\maketitle

\begin{abstract}
 In this paper, we present first-ever optimized hardware implementation of a state-of-the-art neuromorphic approach Histogram of Averaged Time Surfaces (HATS) algorithm to event-based object classification in FPGA for asynchronous time-based image sensors (ATIS).
Our Implementation achieves latency of 3.3 ms for the N-CARS dataset samples and is capable of processing 2.94 Mevts/s.
Speed-up is achieved by using parallelism in the design and multiple Processing Elements can be added. As development platform,  Zynq-7000 SoC from Xilinx is used. The tradeoff between Average Absolute Error and Resource Utilization for fixed precision implementation is analyzed and presented. The proposed FPGA implementation is $\sim$ 32 x power efficient compared to software implementation.
\end{abstract}

\section{Introduction}
Conventional cameras encode the observed scene by producing dense information at a fixed frame-rate which is an inefficient way to encode natural scenes. In bio-inspired neuromorphic event-based cameras \cite{posch2014retinomorphic,posch2011qvga,serrano2013128}, the  individual pixels asynchronously emit events when they observe a sufficient change of the local illuminance intensity. Thus, they have advantages in terms of high temporal resolution, low power consumption, high dynamic range and lower memory requirement. These properties make event-based cameras an ideal choice for remote and embedded applications which are limited in autonomous vehicles, robot navigation or UAV vision. Frame-based computer vision algorithms typically cannot be directly applied to event-based vision data.

The most commonly used architectures for event-based cameras are Spiking Neural Networks (SNN)\cite{cao2015spiking,russell2010optimization,diehl2015fast,kasabov2013dynamic,bohte2002error} which imitate the learning rules observed in biological neural networks. 
While backpropagation is a well-developed general technique for training feedforward deep neural networks, a general technique for training feedforward SNNs is yet to be found.

 Real-time hardware implementations of SNN include IBM’s TrueNorth \cite{haessig2018spiking}, Intel Loihi \cite{davies2018loihi}, SpiNNaker \cite{furber2014spinnaker}, Akida NSoC\cite{akida} and different implementations on FPGA \cite{lammie2018unsupervised,pani2017fpga,neil2014minitaur,cheung2012large}. \\
Histogram of Averaged Time-Surfaces (HATS) \cite{sironi2018hats} is a relatively new event-based machine learning algorithm showing better classification performance compared to Spiking Neural Networks\cite{sironi2018hats}. It relies on a low-level operator called Local Memory Time Surface. It uses local memory units (where neighboring pixels share the same memory block) to efficiently leverage past temporal information and build a robust event-based representation. This representation applies a regularization scheme both in time and space to obtain a compact representation. \\ 
In this paper we implement object-classification using the output from asynchronous time-based image sensors (ATIS)\cite{posch2014retinomorphic}. 
Direct implementation of original HATS on FPGA is not a trivial problem due to high memory requirements to store past events and evaluation of exponential time surface kernels. In this work we show an optimized variant of the HATS suited for FPGA overcoming the above issues.\\
To the best of our knowledge, this is the first FPGA implementation of the HATS algorithm reported in literature. \\
The main contributions of this paper are:
\begin{itemize}
    \item Optimisation of the HATS algorithm for real-world streaming events on FPGA.
    \item Designing a modular implementation using Multiple Processing Elements and Shared BRAM Architecture.
    \item Exploring the trade-off in the fixed-precision implementations of the algorithm on FPGA with Average Absolute Error of the Output. 
\end{itemize}
The remainder of this paper is organized as follows. Section 2 covers the Local Time Surfaces and HATS Algorithm. Section 3 discusses our optimised formulation for inference and dataset used. The FPGA design and implementation is elaborated in Section 4. Section 5 presents the evaluation results of the system and performs a comparison with equivalent software and hardware implementations. Finally, Section 6 includes the conclusions of the paper.

\begin{figure*}[bthp]
\includegraphics[scale=0.7]{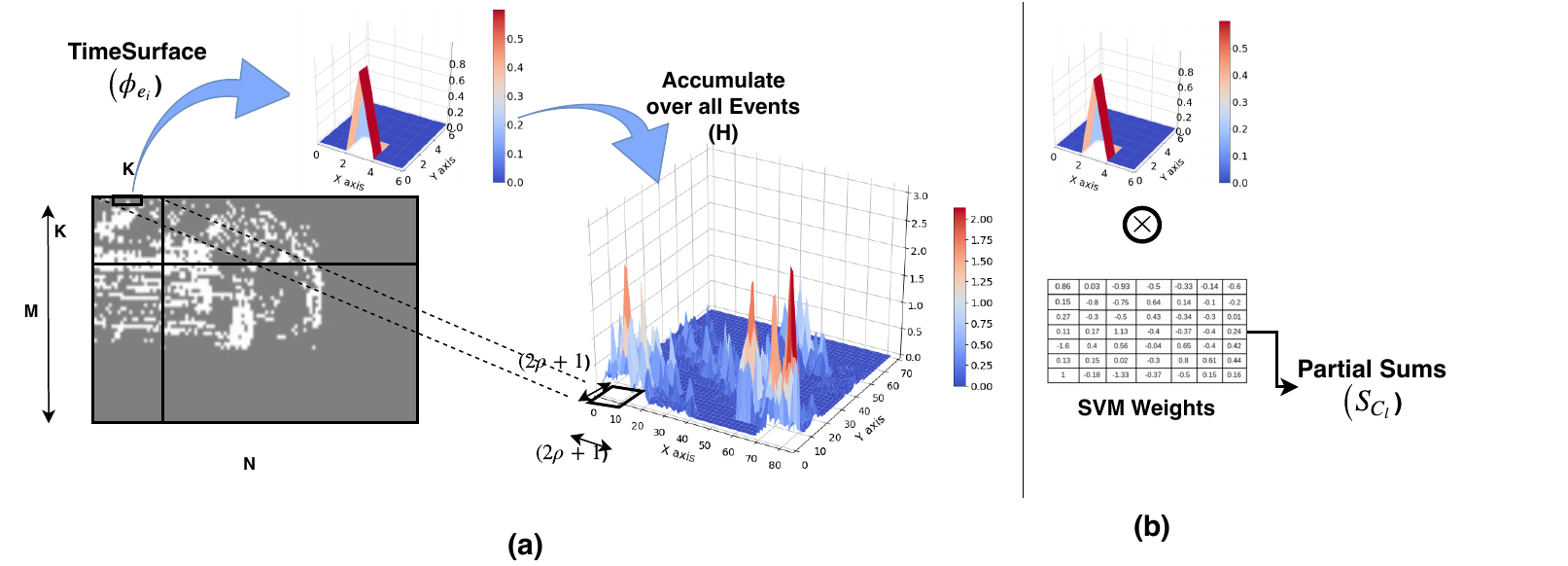}
\centering
\caption{(a) Left Image shows a Car sample of size $M \times N$ from the N-CARS dataset collapsed in time. Right Image: Histogram of Averaged Time Surfaces (HATS) representation of size $(2\rho+1)\lfloor{M/K}\rfloor \times (2\rho+1)\lfloor{N/K}\rfloor$. (b) Figure shows the partial sum accumulation in CWTS from time surfaces.}
\label{fig:hats}
\end{figure*} 

\section{Local Memory Time Surfaces}
A time surface is a spatio-temporal representation of activities around an event relying on the arrival time of events from neighboring pixels.
Given an event-based sensor with pixel grid size $ M \times N$ , a stream of events is given by a sequence
\begin{center}
$ \varepsilon = \{e_{i}\}_{i=1}^I , with \  e_{i} = (\Vec{x_{i}} , t_{i} , p_{i} )$ \\
\end{center}
where $\Vec{x} = (x_{i} , y_{i} ) \in [1, . . . , M ] \times [1, . . . , N ]$ are the coordinates of the pixel generating the event, $t_{i} \geq 0$ the timestamp at which the event was generated, with $t_{i} \leq t_{j}$ for $i < j$, and $p_{i} \in \{-1, 1\}$ the polarity of the event, with $\{-1, 1\}$ meaning respectively OFF and ON events, and I is the number of events.
The adjacent pixels of the frame are grouped in cells ${\{C_{l}\}_{l=1}^L}$ of size $K \times K$. Each cell has a local shared memory associated with it denoted as $Memory_{C_l}$ which stores the events of the last $\Delta t$ interval. Then, for an event ${e_{i}}$ in cell C, Local time surface for the event $\Phi_{e_{i}}$ is given by  
\begin{equation}
\Phi_{e_{i}}\{\Vec{z}\} = \sum_{Q(z,q)} e^{-\frac{t_{i}-t_{j}}{\tau}} 
\end{equation}
where
\begin{center}
${Q(\textbf{z},q)}_{e_{i}} = e_{j} \in M_{C_{l},q}, \ e_{j} \in \{ \Vec{x_{j}} = \Vec{x_{i}}+\Vec{z} , \Vec{z} \in [-\rho,\rho]^2\}$    
\end{center}
where $q$ is the polarity of event $e_j$ such that $q=p_i$, $\rho$ is the spatial window ($\rho<K$) and $\tau$ is a decay factor. The Local Memory Unit of the Cell given by $Memory_{C_{l},q}$ is updated. 
\begin{equation}
Memory_{C_{l},q} \gets Memory_{C_{l},q}\ \text{U} \ e_{i}
\end{equation}
\begin{equation}
  Count_{C_{l}} \gets Count_{C_{l}} + 1   
\end{equation}
The histogram of a given cell is then calculated as the sum of the time surfaces for all the events falling in that particular cell. 
\begin{equation}
h_{C_l}(\Vec{z}, q)=\sum_{e_{i}\in C_l}\Phi_{e_{i}}(\Vec{z}, q)/Count_{C_l}
\end{equation}
It is then normalized by the number of events in that cell. The HATS representation (shown in Fig. \ref{fig:hats}) is then given by the combined histogram from all the cells in equation 3.
\begin{equation}
\textbf{H} =
\begin{pmatrix}
    \begin{pmatrix}
    x_{11}  & \dots  & x_{1n} \\
    x_{21} & \dots & x_{2n} \\
    x_{n1}  & \dots  & x_{nn} \\
    \end{pmatrix}_{\textbf{h}_{C_1}}
   & {\textbf{h}_{C_2}} & \dots \\
    \vdots &  \ddots & \vdots \\
    \dots & \dots  & {\textbf{h}_{C_L}}
\end{pmatrix} 
\end{equation}
This representation is used with a linear SVM Classifier for giving decision boundaries for k-class object detection and classification.
\section{Our Proposed Optimized Algorithm (CWTS) and Dataset}
\subsection{Continuously Weighted Local Time Surfaces (CWTS)}
For inference, we formulate the above algorithm as Continuously Weighted Local Time Surfaces (CWTS). In our formulation the partial sums are stored directly instead to the complete histogram representation which takes more memory requirements. We infer continuously from streaming data to find objects temporally precisely. Figure 2(a) shows the proposed CWTS algorithm with learned weights for k-class object classification for real event-stream data with a linear-decayed time kernel and computation of the partial sums at each iteration. The partial sums can be normalized and accumulated to obtain k-class sums which gives the final class of the object present.
\begin{figure}[htbp]
\includegraphics[scale=0.4]{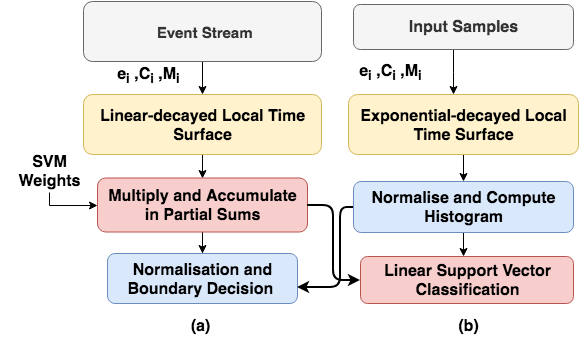}
\centering
\caption{(a) Flow of our Proposed Formulation (b) Top-level Flow of the HATS Algorithm. The cross-arrows show the reversal of order of the blocks from the original implementation.}
\label{fig:data}
\end{figure}


\subsection{Dataset}
We use the N-CARS dataset \cite{sironi2018hats} for our implementation in FPGA.  N-CARS is a challenging dataset which has been acquired in real-world conditions, containing cars at different poses, speeds and occlusions (shown in Fig. \ref{fig:data}). The dataset consists of 7940 car and 7482 background training samples, and 4396 car and 4211 background testing samples. The number of events in each example vary widely from $\sim$ 400 to 18,000 in a 100ms interval. This large variation makes the design of the FPGA implementation difficult.

N-MNIST and N-Caltech101 are also popular datasets which are created by moving the ATIS vision sensor over each image in the popular MNIST, Caltech101\cite{lee2016training}. 
However, Explorations with these datasets have shown that their features encoded are not discriminative in time and do not capture the real world dynamics of the objects\cite{iyer2018neuromorphic}.  \\

\begin{figure}[htbp]
\includegraphics[scale=0.25]{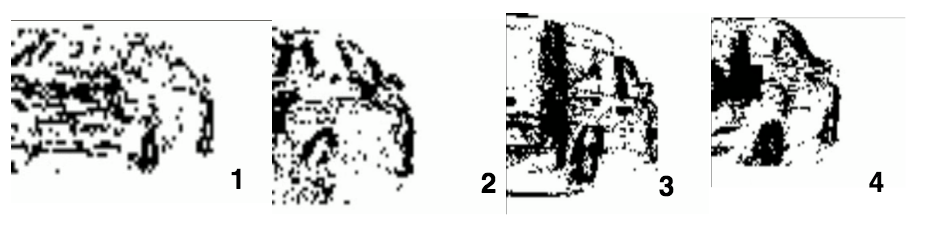}
\centering
\caption{Car images collapsed in time from the N-CARS dataset at different poses, speeds and occlusions.}
\label{fig:data}
\end{figure}

\section{FPGA Implementation}

\begin{table}[!htbp]
  \centering
  \caption{Parameters used in FPGA Implementation}
    \begin{tabular}{|P{1.4cm}|P{0.9cm}|P{1.7cm}|}
    \hline Parameter & Symbol & Value \\
    \hline   Cell Size & K & 10 \\
    \hline  Spatial Window & $\rho$ & 3 \\
    \hline   Temporal Window & $\Delta t$ & 100 ms \\
    \hline   Decay Factor & $\tau$ & $10^6$ms \\
    \hline 
    \end{tabular}%
  \label{tab:parameter}%
\end{table}%

\begin{figure}[htbp]
\includegraphics[scale=0.55]{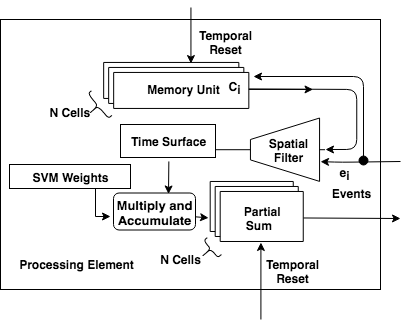}
\centering
\caption{Block Level Flow of the Proposed Processing Element}
\label{fig:pe}
\end{figure}

\begin{figure}[htbp]
\includegraphics[scale=0.5]{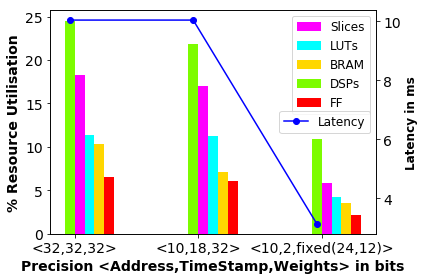}
\centering
\caption{Average Resource Utilisation in PE for different Precision using our proposed implementation in FPGA.}
\label{fig:resource}
\end{figure}

\begin{figure}[htbp]
\includegraphics[scale=0.6]{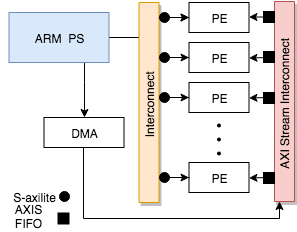}
\centering
\caption{Overall Architecture of the FPGA SoC for the Proposed Implementation.}
\label{fig:block}
\end{figure}

We used Xilinx Zynq xc7z020clg484-1 (Zedboard Evaluation board) SoC FPGA for the HATS implementation. It contains Dual ARM-Cortex A9 Processors (PS7) and Programmable Logic resources. The FPGA implementation directly follows the flow of the Algorithm as shown in Fig. 2(a). The parameters for our FPGA Implementation for a frame size of $120\times100$ are shown in Table \ref{tab:parameter}. \\
In the Processing Element (PE) (shown in Fig. \ref{fig:pe}), the incoming events use the local memory of its corresponding cell. Events in the spatial neighbourhood are then used to calculate the Time surface ($2\rho+1 \times 2\rho+1$) of the event, which is then multiplied with SVM weights to get the local sum. The most used resources are the DSPs where the computation of SVM Weights with Local Time Surface is paralleled to take only 2 iterations (of 11 clock cycles each). The computation is done in fixed-point 24 bits with 12 integer bits. The choice of precision is explained in Section \ref{results}. The SVM weights and Shared Memory Unit of the corresponding cell are stored distributed in BRAMs with access latency of single clock cycle. It has been shown that designing in HLS can achieve upto a 120x speed-up by using the right optimization directives and architecture choices\cite{sotiriou2015evaluation}. We use $ap\_memory$ directive interface to implement external shared BRAMs architecture on FPGA which is allocated for the larger memory requirements. The Local Memory and Local Sum are reset after recieving a temporal reset signal which happens after every $\Delta t$ time period. The design uses microsecond precision timestamps. Using less precise timestamps (10 or 100 microsecond) does not have any effect on accuracy but saves a significant amount of the memory (BRAMs in Fig. \ref{fig:resource}). The BRAM resources can be increased depending upon the application for Shared Memory Unit.   The average resource utilisation and latency decrease from fixed point implementation compared to the full precision implementation is shown in Fig. \ref{fig:resource}. The average latency of each example execution is reduced by $\sim$ 2.5x. \\
Processing Element (PE) were modeled in Vivado HLS (High level Synthesis) and validated using C++/RTL co-simulation and then integration with the ARM-Processor was done using Vivado Design Suite. The resources utilisation when including 8 Processing Element running in parallel is shown in Table \ref{tab:final_resource}. For all the PE we divide the total number of cells equally. The required resources increase linearly with the number of the Processing Elements. Depending on the available FPGA resources, the number of PE can be chosen. Based on the implementation results, the FPGA can operate at a maximum clock speed of 100 MHz. \\
The proposed design is portable and can be implemented on a different SoC-FGPA architecture. The block diagram of the SoC implementation is shown in Fig. \ref{fig:block}. The events are transferred using an AXI-Stream Data FIFO which carries 24-bit AER encoded data. The partial sums are transferred back to the ARM-Processor using the Slave-Axilite Interface.

\begin{table}[!htbp]
  \centering
  \caption{Resource and Power Consumption FPGA Implementation}
    \begin{tabular}{|P{2.1cm}|P{0.7cm}|P{0.8cm}|P{0.8cm}|P{0.6cm}|P{0.7cm}|}
    \hline \textbf{Resources} & \textbf{DSPs} & \textbf{BRAMs} & \textbf{FF} & \textbf{LUTs} & \textbf{Slices}\\
    \hline  \textbf{Total} &  220 & 280 & 106400 & 53200 &13300\\
    \hline \textbf{Utilised} & 192 & 98 & 24038 & 20682 & 8253\\
    \hline \textbf{\% Utilised} &  87.2 & 35 & 22.6 & 33.8 & 62 \\
    \hline \textbf{Power Estimation (mW)} & 169 & 14 & -- & -- & 98\\
    \hline 
    \end{tabular}%
  \label{tab:final_resource}%
\end{table}

\section{Results and Discussions}
\label{results}
Average Absolute Error (AAE) metric was evaluated for a full precision software version of the model and compared to the fixed precision FPGA implementation. To evaluate the performance of the implementation we evaluate average absolute error of the partial sum $Sum_{j}$ on training samples of N-CARS Dataset (shown in Table \ref{tab:precisioni}). When reducing the fractional bits for doing the computation from 12 bits to 7 bits the AAE increases exponentially. We have chosen precision to be $<24,12>$ (12 Integer with 12 Fractional bits) for our implementation which tested on N-CARS Samples does not have any accuracy loss.
The average FPGA latency for the samples in N-CARS positive training samples $\sim$ 3.3 ms/sample. It can process on avg.  $\sim$ 2.94 Mevents/s. A further speedup can be achieved for applications where the spatial window close to the cell size $(\rho \approx K)$, which will allow a much faster parallel execution of the time surface computation. The performance of the Parallelized Python implementation is determined by measuring the computation time required for analysing training images on a PC having an Intel Core i7-8700K CPU, 32 GB DDR4 RAM, Linux Ubuntu 16.04 which takes 34.4 ms/sample. Python implementation is used for training and testing the algorithm for this dataset which gives an accuracy of 87.5 \%.
To make the comparison possible both implementations of the hardware and software use exactly the same parameters. The total power consumptions of the SoC FPGA implementation is $\sim$ 0.62 W (estimated in Vivado Design Suite and Xilinx Power Estimator), ARM PS is $\sim$ 0.17 W (at 222 MHz) and programmable logic is $\sim$ 0.3 W and static power consumption is 0.15W. The estimated power consumption  done using the tool powerstat\cite{powerstat} on a PC with Intel i5-6300U, 12 DDR4 RAM, Linux is $\sim$ 19.2 W. The SoC FPGA implementation is $\sim$ 32 x power efficient.
\begin{table}[!htbp]
  \centering
  \caption{Precision of Computation  vs Average Absolute error }
    \begin{tabular}{|P{2.4cm}|P{1.4cm}|}
    \hline \textbf{Precision fixed $<total,integer> in bits$} &  \textbf{AAE (\%)} \\ 
    \hline $<24,12>$  & 1.4 \\
    \hline $<23,12>$  & 2.9 \\
    \hline  $<22,12>$  & 5.8 \\
    \hline  $<21,12>$  & 11.5 \\
    \hline  $<20,12>$  & 23.3 \\
    \hline  $<19,12>$  & 45.3 \\
     \hline 
    \end{tabular}%
  \label{tab:precisioni}%
\end{table}%
\section{Conclusions}
We have presented first ever reported implementation of the Histogram of Averaged Time Surfaces (HATS) algorithm for event-based object classification for asynchronous time-based image sensor (ATIS) in FGPA. The proposed architecture is flexible, as it allows the use of multiple instances of the Processing Elements. This makes developers free to choose the speed range and reserved resources for this task. The design is capable of comfortably processing 2.94 Mevts/sec. 
The proposed implementation achieves lower latency/sample compared to the software version and the 24-bit fixed precision FPGA algorithm performs very similarly to the full precision software version. \cite{radway2021future, bashir2019power, sethi2021efficient, ji2020reconfigurable,ji2020compacc, sethi2020nv, sethi2020design, sethi2018low, sethi2019optimized, sethi2022dragon}
\section*{Acknowledgements}
This research activity under the PI Prof. M. Suri is supported by the IIT-D FIRP grants. 

\bibliographystyle{IEEEtran}
\bibliography{ref}

\end{document}